\documentclass[twocolumn,a4paper,preprintnumbers,amsmath,amssymb,floatfix,prl]{revtex4}

\usepackage{graphicx}% Include figure files
\newcommand{\bm}[1]{\mbox{\boldmath$#1$}}

\begin{document}

\title{Quantum State of Neutrons in Magnetic Thin Films
and Superlattices}

\author{Florin Radu$^{1,2}$}
\email{florin.radu@bessy.de}
\homepage{http://www.ep4.ruhr-uni-bochum.de}
\author{Hartmut Zabel$^1$}
\affiliation{$^1$Department of Physics, Ruhr-University Bochum,
44780 Bochum, Germany} \affiliation{$^2$BESSY GmbH,
Albert-Einstein-Str. 15, 12489 Berlin, Germany}

\begin{abstract}

An experiment which describes the quantum states of neutrons in
magnetic thin films and superlattices is reviewed.

\end{abstract}
%\pacs{74.78.Fk, 61.12.Ex,  75.25.+z} %\maketitle

\date{\today}
%\keywords{Magnetic multilayers, polarized neutron scattering}
\maketitle

%%      Your contribution follows:
%%      DO NOT USE subsection and subsubsection

%\section{Introduction}
While specular polarized neutron reflectivity (PNR) is widely
recognized as a powerful tool for the investigation of
magnetization profiles in magnetic
nanostructures~\cite{Fitzsimmons}, there is still a confusion
concerning the quantum state of neutrons in a magnetic sample. We
have now shown unambiguously that the neutron has to be treated as
a spin 1/2 particle in each homogeneous magnetic layer. This is at
variance with the conventional description of neutron
reflectivity, which often considers the neutron magnetic potential
as a classical dot product.

Neutrons interact with a magnetic thin film via the Fermi nuclear
potential and via the magnetic induction. Thus, the neutron - film
interaction hamiltonian includes both contributions:
$V=V_n+V_m=(\hbar^2/2m) 4\pi N b-{\bm{\mu B}}$, where $m$ is the
neutron mass, $N$ is the particle density of the material, $b$ is
the coherent scattering length,  $|{\bm{\mu}}|$ is the magnetic
moment of the neutron, and $|{\bm{B}}|$ is the magnetic induction
of the film. Unconventionally, however, neutron reflectivity
treats the dot product between the magnetic induction and neutron
magnetic moment classically: $V_m=-\bm{\mu \bullet B}=\pm
\bm{|\mu| |B|}\cos(\theta)$, where $\theta$ is the angle between
the incoming neutron polarization direction and  the direction of
the magnetization inside the film. Writing the magnetic potential
as a classical dot product implies that the neutron energies in
the magnetic layer have a continuous distribution from
-$|\bm{\mu}| |\bm{B}|$ to +$|\bm{\mu}| |\bm{B}|$. This predicts
that the critical angle for total reflection depends on the angle
between the direction of polarization and the direction of the
magnetic field inside the layer.

On the other hand we know that there are only two eigenstates for
the spin 1/2 particles in a magnetic field. Therefore, the eigen
wave number of a neutron in a magnetic thin film has two proper
values. After solving the Schr\"odinger equation one obtains two
eigen wave numbers for neutrons in a magnetic film.   It follows
that there are only two possible energies and consequently only
two values for the index of refraction corresponding to the
spin-up and spin-down states of the neutrons. Therefore, quantum
mechanics predicts that there are only two critical angles for the
total reflection: one corresponding to the R$^+$ and one to the
R$^-$ reflectivity.

\begin{figure}[!h]
\begin{center}
\includegraphics[clip=true,keepaspectratio=true,width=1\linewidth]{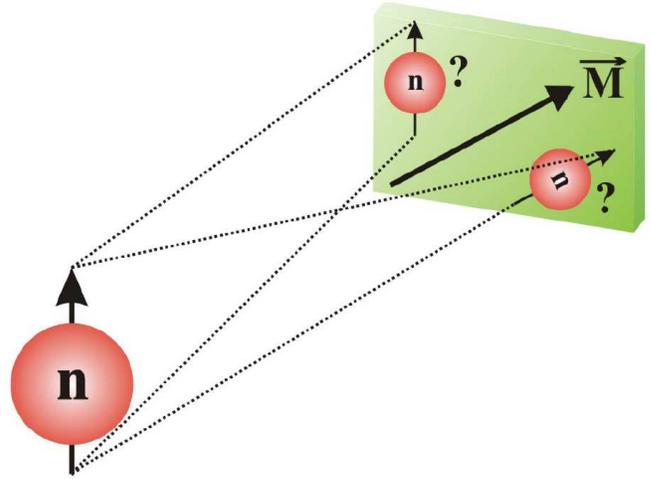}
\end{center}
 \caption{\label{fig1} Two possible orientations of
the neutron spin in a film with a homogeneous magnetization M,
which makes an angle $\theta$ against the polarization axis:
either the spin orientation is maintained parallel to the
polarization axis, or the spin is reoriented according to the new
quantization axis in the film. Maintaining the spin orientation
corresponds to a classical description of the interaction of
neutrons with a magnetic potential, reorientation is required by
quantum mechanics for a particle with spin 1/2. }
\end{figure}

Obviously there is a contradiction between the quantum mechanical
prediction and the prediction based on the classical
representation of the magnetic potential: quantum mechanics
predicts that the spin state of the neutron is determined by the
magnetic induction in the sample, whereas classical representation
of the magnetic potential, supported by experiments on magnetic
multilayers, assert that the spin state of the neutrons is fixed
by the incident polarization axis (see Fig.~\ref{fig1}).

To resolve this contradiction we have carried out an experiment
which provides direct and unambiguous evidence for the spin state
of neutrons in magnetic media. The goal was to find a system where
the angle between the neutron polarization and direction of the
magnetization inside of the film can be fixed and controlled. Then
we measure the R$^+$ and R$^-$ reflectivities and determine
whether the position of the critical edges changes as a function
of the angle , or whether the critical edges stay fixed, and only
intensity redistributes between reflections R$^+$  and R$^-$ with
change of $\theta$ . The easiest way to control the angle $\theta$
is to rotate the magnetic film and therefore the magnetization
direction with respect to the neutron spin polarization, which
remains fixed in space outside of the sample. This requires that
the film should have a high remanent magnetization. Additionally,
the film thickness should exceed the average neutron penetration
depth. The last requirement is essential in order to avoid neutron
tunnelling effects which will hinder the precise determination of
the critical edges.

\begin{figure}[!h]
\begin{center}
\includegraphics[clip=true,keepaspectratio=true,width=1\linewidth]{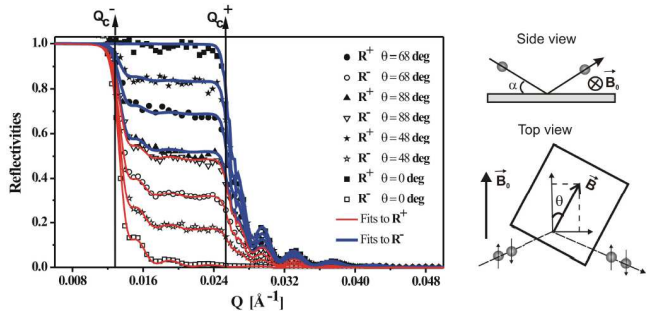}
\end{center}
\caption{\label{fig2} Experimental results of reflectivity curves
R$^+$ and R$^-$ from 100 nm thick Fe film on a Si substrate. The
reflectivities are plotted on a linear scale. The two sets of
R$^+$ (solid black symbols) and R$^-$ (open black symbols)
reflectivity curves were measured for four different angles
between the neutron polarization and the film magnetization vector
(i.e., the magnetic induction B in the sample plane). The blue and
red lines are the simulated R$^+$ and R$^-$ reflectivities,
respectively. In panels on the right side the experimental
geometry are shown. The experiments show that the critical edges
$Q_c^{+}$ and $Q_c^{-}$ do not depend on the $\theta$ angle.}
\end{figure}

 The experiment was performed at the ADAM reflectometer
(ILL). As magnetic film we have chosen a 100 nm thick Fe layer
deposited by rf-sputtering on a Si substrate. The remanence
obtained by measuring the hysteresis curve is about 97.5\%. In the
rotation experiment the Fe layer was first magnetized parallel to
the neutron polarization direction and then the magnet was
removed. A small guiding field was still present at the sample
position in order to maintain the neutron polarization.
Subsequently a series of R$^+$ and R$^-$ reflectivities were
measured for several in-plane rotation angles of the sample. The
results are shown in Fig.~\ref{fig2}. We observe two
characteristics of the reflectivities: first, the critical edges
are fixed and independent of the in-plane rotation angle $\theta$
of the magnetization vector, and second the R$^-$ intensity
continuously increases at the expense of the R$^+$ intensity as a
function of the  $\theta$ angle. The plain experimental results as
well as a detailed fit unambiguously show that the critical edges
$Q_c^+$ and $Q_c^-$ for total reflection of the two spin states
are fixed independent of the orientation of the magnetization
vector in the film. This confirms that the neutron spin inside of
the sample is oriented according to the quantization axis parallel
or antiparallel to the magnetization vector.

\begin{figure}[!h]
    \begin{center}
      \includegraphics[clip=true,keepaspectratio=true,width=1\linewidth]{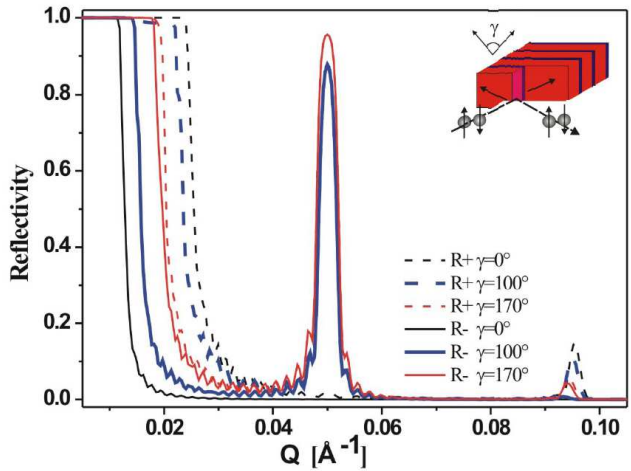}
      \caption{Simulation of polarized neutron reflectivities
(R$^+$ and R$^-$) for a Fe/Cr multilayer as a function of coupling
angle ($\gamma$) between the magnetization vectors of adjacent Fe
layers. In the inset the superlattice is shown with a cut-away
view of the top magnetic layer. The critical angle for total
reflection is sensitive to the coupling angle ($\gamma$) but not
to the orientation of the average magnetization, with respect to
the polarization axis of the neutrons. Also shown are the
halforder peak from the antiferromagnetic component of the
superlattice at $Q \approx 0.05 \AA^{-1}$ and the first order
ferromagnetic peak at $Q \approx 0.095 \AA^{-1}$.}
      \label{fig3}
    \end{center}
\end{figure}

So far so good for magnetic thin films. But what is the
orientation of polarized neutrons in magnetic multilayers with
periodically varying magnetization direction? Does the spin
orientation of the neutron follow the rapid change from layer to
layer, or does the neutron interact with an average potential? To
answer this question we have simulated the reflectivity profile of
a Fe(6 nm)/Cr(0.8 nm) superlattice containing 40 repeats. The
thicknesses of the Fe and Cr layers chosen are typical for this
class of materials.  For the simulation we used the freeware code
PolarSim,  which is based on the generalized matrix
method~\cite{polar}. In Fig.~\ref{fig3} simulations are shown for
R$^+$ and R$^-$ reflectivities and for three angles $\gamma$
between the magnetization vectors of adjacent Fe films: 0, 100°,
and 170° , the last one being close to an antiferromagnetic
orientation of the Fe layer. Our focus is on the behaviour of the
critical scattering vector for total reflection. We observe that
for $\gamma$ = 0° the $Q_c^+$ and $Q_c^-$ are well separated and
that they contain information about the saturation magnetization.
When the   value increases, the critical edges approach each
other. For an angle $\gamma$ = 180° (not shown here) there is no
difference between the R$^+$ and R$^-$ reflectivities in the
vicinity of the critical scattering vector. The main result from
this simulation is the observation that the separation of the
critical edges is a continuous function of the angle   between the
in-plane adjacent magnetization vectors. Thus, unlike the single
magnetic film, in case of magnetic multilayers the critical edges
move. However, they do not shift according to the angle between
the average magnetization vector and the polarization axis (as
expected in the classical case), but according to the angle
between the magnetization vectors between adjacent magnetic
layers. Thus, from a practical point of view it is sufficient to
determine the difference between the critical edges  $Q^2 =
(Q^+_c)^2 - (Q^-_c)^2$ to measure the angle between the
magnetization vectors in magnetic superlattices, which than
follows from: $\cos( \gamma /2)=\Delta Q^2 (\hbar^2/2m) /
(2|\bm{\mu}| |\bm{B}|)$.

If we now choose a fixed coupling angle $\gamma$  between the
magnetization vectors of adjacent layers and rotate the sample as
in the previous case of a single magnetic layer, then the critical
edges behave a again in accordance with the neutron spin states in
homogeneous magnetic media, i.e. $Q_c^+$ and $Q_c^-$  remain
constant, but the intensity varies with the rotation angle between
the average magnetization vector and the polarization direction.

This finally solves a long lasting controversy about the
dependence of the critical edges for total reflection of polarized
neutrons. The apparent shift of the critical edges is solely due
to the opening angle between the magnetization vectors in
successive layers of a magnetic multilayer. As soon as this angle
is fixed anywhere between a ferro- to an antiferromagnetic state,
the critical angles are fixed independent of the orientation of
the superlattice with respect to the polarization direction of the
neutrons. In the single homogeneous magnetic layer as well as in
the superlattice, the polarized neutron within the sample assumes
new eigenstates accordance with the average magnetic induction in
the sample.

This work was supported by the Deutsche Forschungsgemeinschaft
through the research network (SFB 491): Magnetic heterostructures,
which is gratefully acknowledged. The ADAM reflectometer is
supported by BMBF contract O3ZA6BC1.


\begin{thebibliography}{99}
\bibitem{Fitzsimmons} M.R. Fitzsimmons , S. D. Bader, J. A. Borchers, G. P. Felcher, J. K. Furdyna,
A. Hoffmannb, J.B. Kortrighte, Ivan K. Schullerf, T.C. Schulthess,
S. K. Sinha, M. F. Toney, D. Weller,  S. Wolf, J. Magn. Magn.
Mater. 271, 103 (2004).
\bibitem{polar}
Polarized Neutron Reflectivity Software
(\textit{PolarSim}),http://www.ep4.rub.de/\~radu/welcome/polar.html
\bibitem{fr}
Full reference to the work described here: F. Radu, V. Leiner, M.
Wolff, V. K. Ignatovich and H. Zabel. Quantum state of neutrons in
magnetic thin films. In \textit{Physical Review B} \textbf{71},
214423 (2005).
\end{thebibliography}
\end{document}